\documentclass[sigconf]{acmart}

\copyrightyear{2025}
\acmYear{2025}
\setcopyright{acmlicensed}\acmConference[KDD '25]{Proceedings of the 31st ACM SIGKDD Conference on Knowledge Discovery and Data Mining V.2}{August 3--7, 2025}{Toronto, ON, Canada}
\acmBooktitle{Proceedings of the 31st ACM SIGKDD Conference on Knowledge Discovery and Data Mining V.2 (KDD '25), August 3--7, 2025, Toronto, ON, Canada}
\acmDOI{10.1145/3711896.3737232}
\acmISBN{979-8-4007-1454-2/2025/08}

\usepackage[linesnumbered,ruled,vlined]{algorithm2e}
\usepackage{subcaption}
\usepackage{amsmath}
\usepackage{balance}

\DeclareMathAlphabet{\mathbbold}{U}{bbold}{m}{n}

\begin{document}

\title{Harnessing the Power of Interleaving and Counterfactual Evaluation for Airbnb Search Ranking}

\author{Qing Zhang}
\orcid{0009-0005-3572-2606}
\affiliation{%
  \institution{Airbnb}
  \city{San Francisco}
  \state{CA}
  \country{USA}
}
\email{qing.zhang@airbnb.com}

\author{Alex Deng}
\authornote{Work completed while employed by Airbnb}
\affiliation{%
  \institution{Microsoft}
  \city{Seattle}
  \state{WA}
  \country{USA}
}
\email{alexdeng@live.com}

\author{Michelle Du}
\affiliation{%
  \institution{Airbnb}
  \city{San Francisco}
  \state{CA}
  \country{USA}
}
\email{michelle.du@airbnb.com}

\author{Huiji Gao}
\affiliation{%
  \institution{Airbnb}
  \city{San Francisco}
  \state{CA}
  \country{USA}
}
\email{huiji.gao@airbnb.com}

\author{Liwei He}
\affiliation{%
  \institution{Airbnb}
  \city{Seattle}
  \state{WA}
  \country{USA}
}
\email{liwei.he@airbnb.com}

\author{Sanjeev Katariya}
\affiliation{%
  \institution{Airbnb}
  \city{San Francisco}
  \state{CA}
  \country{USA}
}
\email{sanjeev.katariya@airbnb.com}

\renewcommand{\shortauthors}{Qing Zhang et al.}

\begin{abstract}
Evaluation plays a crucial role in the development of ranking algorithms on search and recommender systems. It enables online platforms to create user-friendly features that drive commercial success in a steady and effective manner. The online environment is particularly conducive to applying causal inference techniques, such as randomized controlled experiments (known as A/B test), which are often more challenging to implement in fields like medicine and public policy. However, businesses face unique challenges when it comes to effective A/B test. Specifically, achieving sufficient statistical power for conversion-based metrics can be time-consuming, especially for significant purchases like booking accommodations. While offline evaluations are quicker and more cost-effective, they often lack accuracy and are inadequate for selecting candidates for A/B test.
To address these challenges, we developed interleaving and counterfactual evaluation methods to facilitate rapid online assessments for identifying the most promising candidates for A/B tests. Our approach not only increased the sensitivity of experiments by a factor of up to 100 (depending on the approach and metrics) compared to traditional A/B testing but also streamlined the experimental process. The practical insights gained from usage in production can also benefit organizations with similar interests.
\end{abstract}

\ccsdesc[500]{Mathematics of computing~Hypothesis testing and confidence interval computation}
\ccsdesc[500]{Applied computing~Electronic commerce}

\keywords{causal inference, interleaving, counterfactual evaluation, search ranking, recommendation}

\title{Harnessing the Power of Interleaving and Counterfactual Evaluation for Airbnb Search Ranking}
\maketitle

\section{Introduction}
\label{sec:intro}
Web platforms extensively utilize data-driven approaches to refine their ranking algorithms for search or recommendation systems. By presenting users with different variations, we can measure the impact of changes based on user actions directly. A/B testing is a widely used method for this purpose. In a typical A/B testing scenario, visitors to a website are randomly assigned to either a control group or a treatment group. The control group is exposed to the baseline version of the website, which usually reflects the current live production environment, while the treatment group experiences the versions produced by the new, proposed algorithm. The effectiveness of the treatment is evaluated by comparing key business metrics, such as conversion rates, between the two groups. Additionally, a comprehensive set of debugging metrics, including funnel conversion rates, user engagement levels, and characteristics of the results, are analyzed to gain deeper insights into the behavior and performance of the ranking algorithms. A/B testing has become a cornerstone for fostering continuous innovations, playing a crucial role in their ability to adapt and improve \cite{kohavi2020trustworthy}.

A significant volume of work has been dedicated to enhancing the effectiveness of A/B testing. Notably, \cite{deng2013improving} proposed leveraging pre-experiment data to reduce metric variance and improve sensitivity. This approach has been widely adopted across the industry, as seen in works such as \cite{xie2016improving, liou2020variance}. Additionally, in-experiment data has been utilized to develop surrogate metrics that aim to achieve high sensitivity and provide early readings on business metrics, which often experience delayed outcomes \cite{deng2023variance}. The extensive use of A/B testing in the industry has led to the accumulation of valuable practical lessons, which have been studied in depth \cite{kohavi2009controlled, kohavi2024false, dmitriev2016pitfalls}. 

A/B testing alone, however, is often insufficient due to the long running times required for experiments on e-commerce platforms like Airbnb. There are three main reasons. First, users typically visit the site with a certain level of intent to make a purchase. Since Airbnb users generally travel only twice a year, the traffic for experiments is significantly lower compared to search engines. Second, while conversion is usually the target metric, it takes time to realize. The higher the stakes, the longer the search journey; for instance, an Airbnb user might spend several days searching for accommodation that fits their preferences. In contrast, search engines receive immediate feedback from user clicks. Third, as the system matures, the effect size of each innovation tends to be small, which necessitates an even longer time to detect statistically significant changes \cite{bi2022debiased}.

A logical question arises: why don't we use offline evaluations to identify the most promising candidates for A/B testing? Indeed, offline evaluation is often employed as a preliminary step for model assessment. This process involves collecting search logs, which include the results displayed and the corresponding user actions, and using them to evaluate the proposed ranker. While this approach is efficient and risk-free, it tends to lack accuracy. The primary reason is that the ranker we aim to evaluate only has the visibility of what has been shown by the logging ranker but not all the candidate items. To address this selection bias, techniques such as inverse propensity weighting (IPW) \cite{rosenbaum1983central}, based on importance sampling \cite{horvitz1952generalization}, have been proposed. However, these techniques often result in high variance, as noted by \cite{gilotte2018offline}. In addition, obtaining the propensity score (the probability of a result being displayed) is complicated due to system complexity \cite{chen2019top}.

In addition, offline metrics are frequently disconnected from online business metrics. For example, the Normalized Discounted Cumulative Gain (NDCG) \cite{jarvelin2002cumulated} is a standard offline metric, while conversion rate serves as the primary online metric. Often, these two metrics are inconsistent. Furthermore, offline evaluations cannot fully account for the user dynamics that occur when individuals interact with ranked lists.

To bridge the gap between the two approaches, we seek a middle step that is faster than A/B testing while being more accurate than offline evaluation. Interleaving experiments, first proposed in \cite{joachims2002optimizing, joachims2003evaluating}, offer a potential solution. However, several open questions remain. Firstly, the interleaving methods developed in prior work primarily used clicks as user feedback, whereas our target metric is conversion, which is much sparser, as previously noted. Additionally, we must consider scalability, as the method will be implemented in production and handle real traffic; thus, both complexity and latency need to remain within acceptable limits. Consequently, it is unclear whether the state-of-the-art interleaving mechanisms can be efficiently and effectively applied to ranking problems in e-commerce platforms like Airbnb.

In this paper, we share our recent advances in enhancing the experimentation velocity for Airbnb's search ranking. We present our innovations in interleaving experiment design and the engineering framework. Following this, we detail an online counterfactual evaluation approach that is more generalizable and addresses the limitations of interleaving. Both techniques are utilized for selecting treatment candidates for A/B testing. These systems are used by engineers working on search ranking at Airbnb, and the validation of these techniques is based entirely on real-world usage, as opposed to the dataset-based simulations commonly used in previous work. Our contributions are, 

\begin{itemize}
\item Competitive pair based interleaving that's unbiased and highly efficient, and we observed 50X speedup improvement compared to A/B in production. The speedup is computed with traffic needed to achieve similar statistical power as A/B test.
\item Online counterfactual evaluation which further improved the sensitivity on top of interleaving and more generalizable. The metrics demonstrated up to 100X speedup compared to A/B.
\item Practical lessons from the usage in production that provides full picture of of the techniques.
\item Both interleaving and counterfactual evaluation approaches presented in the paper can be fairly easily generalized to other platforms.
\end{itemize}

To the best of our knowledge, this is the first work where both approaches have been implemented in production, evaluated side by side and used on a daily basis. The paper provides a comprehensive comparison, detailing both the advancements and limitations of each approach and an experimentation strategy with interleaving, counterfactual evaluation, and A/B test to improve overall experimentation velocity in practice, which will prove invaluable for businesses facing similar challenges.

\section{Preliminaries}
\subsection{Problem Definition}
\label{sec:problem_def}
First, we define the notations and formulate the evaluation problem. We use $\pi$ to denote the ranking algorithm (also referred to as the policy). Given a set of candidate listings with features $X$ (which include listing attributes, past engagements, queries, and user history), the algorithm generates a ranked list represented as $L \sim \pi(L | X)$. After presenting this list to the user, we observe the reward $O \sim p(O | X, L)$, which can include events such as clicks and purchases. $O$ can be an empty list when there is no user action for the $L$. The value of the policy $\pi$ is defined as the expected reward $V(\pi) = E(f(O) * {\pi}(O | X))$, where $f$ maps rewards into numeric value. Given a proposed policy $\pi_1$ and a baseline policy $\pi_0$, the evaluation problem involves designing an estimator for $V(\pi)$ and using the difference $\tau = V(\pi_1) - V(\pi_0)$ to assess the impact of policy $\pi_1$ compared to policy $\pi_0$.

The intuition of comparison can be further developed with potential outcome framework \cite{imbens2015causal}. Formally, let $W$ be the assignments, and each element $w_i$ indicates the group that subject $i$ belongs to, specifically control when $w_i=0$ and treatment when $w_i=1$. Also, let $Y$ denote the outcome (reward), where $Y_{i}(0)$ represents the outcome when control is applied to instance $i$, and similarly $Y_{i}(1)$ treatment. For a moment, let's assume we can observe outcomes from both groups then we can compute the impact $ Y_{i}(1) - Y_{i}(0)$ on each element, and get average treatment effect

\begin{equation}
\label{eq:potential_outcome}
\tau = \frac{1}{N}\sum_{i=1}^{N} (Y_{i}(1) - Y_{i}(0)) = \bar{Y}(1) - \bar{Y}(0)
\end{equation}
Therefore $V(\pi_w)=\bar{Y}(w)$ here. In common settings we cannot observe both $Y_{i}(0)$ and $Y_{i}(1)$ at the same time. For example a user is either exposed to the treatment or control policy, but not both. Thanks to the randomized controlled experiment, such as A/B test, the difference between the observed outcome $\hat{\tau}$ is unbiased for $\tau$ \cite{imbens2015causal},
\begin{equation}
\hat{\tau} = \bar{Y}_t^{obs} - \bar{Y}_c^{obs}
\end{equation}
where superscript $obs$ indicates that these are observed outcomes.
$\bar{Y}_c^{obs}=\frac{1}{N_c}\sum_{i:w_{i} = 0}Y_{i}^{obs}$, and similarly $\bar{Y}_t^{obs}=\frac{1}{N_t}\sum_{i:w_{i} = 1}Y_{i}^{obs}$. 

Interleaving and counterfactual evaluation, on the other hand, connects with the original form Eq ~\ref{eq:potential_outcome} by examining the $Y_{i}(1) - Y_{i}(0)$. We will have more discussion in later sections.

Percent delta relative to baseline is computed as following, 
\begin{equation}
\label{eq:delta}
\%\Delta = \hat{\tau} / \bar{Y}_c^{obs}
\end{equation}
In rest of the paper we omit the step of Eq ~\ref{eq:delta} when discussing metrics for simplicity and assume it is always the final step.

\subsection{Design Principles}
When examining and designing the online evaluation techniques, we are guided by the following principles that were proposed and refined in previous work \cite{radlinski2013optimized, hofmann2013fidelity,joachims2003evaluating,bi2022debiased},
\begin{itemize}
\item {\bf Sensitivity}. A primary objective of an evaluation approach is to achieve the desired statistical power with the minimal amount of data necessary.
\item {\bf Unbiasness}. When a user randomly interact with ranked result, we should expect there is no preference according to the estimator.
\item {\bf Fidelity}. The estimator should align with the intuition when user operate on the original results.
\item {\bf Minimal user experience disruption}. The user experience during the evaluation should mirror the experience that they would typically have when using the product under normal conditions.
\item {\bf Acceptable complexity and scalability}. The system is going to be integrated into a large-scale user-facing search framework, necessitating efficiency in both operation and maintenance.
\item {\bf Generalizability}. Given the dynamic nature of the search system and evolving business requirements, the evaluation methodology must be flexible and easily extendable.
\end{itemize}
It is expected that an evaluation approach may perform well in some criteria while underperforming in others. Therefore, trade-offs have to be made based on the specific use case.

\section{Related Work}

\subsection{Interleaving Experiments}
\label{sec:bg_iterleaving}
Interleaving is an online testing methodology first proposed in \cite{joachims2002optimizing, joachims2003evaluating}. The central idea is to provide the same user with two variants that we want to compare (e.g., results from two rankers) and to infer their preferences based on the user actions, which indicate the quality difference between the two. The key components of this methodology include a merging algorithm to blend the two result lists and a credit assignment mechanism. The technique enables the assessment of ranker relevance through user events, such as clicks, without adding any overhead for the user. Moreover, it allows for a direct comparison of two rankers by the same user. This development is significant, as it marked a shift away from the traditional reliance on human annotators for Web search evaluation.

There are primarily three types of interleaving methods. The first type, Balanced Interleaving (BI), was introduced in \cite{joachims2002optimizing, joachims2003evaluating}. This method ensures that the top $k$ results in the merged result list $I$ consistently include the top $k_a$ results from ranker A and $k_b$ results from ranker B, with $k_a$ and $k_b$ differing by no more than one \cite{chapelle2012large}. As a result, the merged list evenly distributes impressions between the two rankers. For credit assignment, each click within $I$ is attributed to both A and B, provided it appears in their respective lists and is above a certain position threshold. The ranker accumulating more clicks is deemed superior. However, as \cite{chapelle2012large} points out, balanced interleaving can yield biased results when the two rank lists are almost identical, differing only by a slight shift or insertion. To address this bias, \cite{bi2022debiased} developed an interleaving method that debiases BI by incorporating IPW for credit attribution. The process requires, for each ranker, computing the probability of receiving a click at each position in the merged list. Despite its effectiveness, the BI+IPW method poses complexities in credit attribution and less extensible compared to our approach.

The second type of interleaving method is Team Drafting Interleaving (TD), as introduced in \cite{radlinski2008does}. This method utilizes a merging algorithm that mimics the process of drafting in sports teams. It iterates through both ranked lists from top to bottom, selecting the highest-ranked available item to add to the combined list $I$. Each item in $I$ is assigned to a "team," indicating its origin from either ranker. Preferences are determined by counting which "team" has gathered more clicks. This approach addresses the bias issue identified with BI. \cite{chapelle2012large} suggested several refined schemes for credit assignment. Our work builds upon TD, offering more efficient credit computation and increased generalizability. \cite{hofmann2013fidelity} pointed out that TD potentially violates fidelity in certain cases, our practical experience in production has not revealed any issue stemming from these cases. Moreover, our research into counterfactual evaluation indicates that such scenarios have a negligible effect on the overall evaluation. We will explore this topic in greater detail in Section ~\ref{sec:connection_with_interleaving}.

The third type of interleaving method, Probabilistic Interleaving (PI), was introduced in \cite{hofmann2011probabilistic,hofmann2013fidelity} with the aim of improving upon BI and TD. Unlike BI and TD, where the merged list $I$ is constructed from a fixed order of items from A and B, PI uses softmax functions s(A) and s(B) to transform these lists into probability distributions over documents, from which items are then sampled to create list $I$. The credit computation in PI considers all possible sequences of drafting that could result in the formation of list $I$. While PI is unbiased in its approach, it has the potential to significantly alter the user experience and introduces greater system complexity for production use. 

\subsection{Interleaving in Practice}
There is limited literature on how interleaving is used in production. Most of the previous work are research projects ran on limited datasets, such as \cite{hofmann2011probabilistic, hofmann2013fidelity}. There were experiments conducted in Microsoft and Yahoo!, such as \cite{chapelle2012large} but consistency with A/B test were not discussed. The study most relevant to our work is \cite{bi2022debiased} by Amazon, which uses BI as base algorithm and applied IPW to correct the bias, as mentioned earlier. It was evaluated by comparing the results with 10 A/B tests. Our approach is more efficient and we report our comparison with A/B with much larger corpus.

\subsection{Counterfactual Evaluation}
When evaluating policy $\pi$, data collected from another policy $\pi_0$ is often used. It is usually in the form of past search logs, leading to the terms "off-policy" and "offline evaluation" being used interchangeably. Three categories of work have emerged in this field. The first category involves directly modeling (DM) the reward and using it to predict the outcome of the target policy for each search. This approach typically results in a low variance but high bias estimator \cite{joachims2016counterfactual}.

The second category is the model-free approach. As mentioned in Section ~\ref{sec:intro}, Inverse Propensity Weighting (IPW) method is proposed to correct the probability of events observed in historical data. For instance, if an item has a probability $\pi$ of being shown according to the target policy, the outcome is weighted by $ \frac{\pi}{\pi_0}$, where $\pi_0$ represents the probability of being shown according to the logging policy \cite{rosenbaum1983central, horvitz1952generalization}. While IPW is unbiased, it suffers from high variance.  A series of techniques have been developed to address the challenges in IPW based approaches, including Clipped Inverse Propensity Score \cite{swaminathan2015batch} and Self-Normalized IPS estimator \cite{swaminathan2015self}.

In the third category, the Doubly Robust Estimator \cite{dudik2011doubly} combines the DM and IPW estimators. This approach is both unbiased and consistent while exhibiting lower variance than IPS. Furthermore, several variations have been developed based on the doubly robust estimator \cite{su2020doubly, su2019cab, wang2017optimal, metelli2021subgaussian}.

The aforementioned work primarily focuses on multi-arm bandit evaluation, which cannot be directly applied to search ranking due to the large action space.  However, in the online counterfactual evaluation discussed in this paper, we incorporate elements of IPW and reward estimation. 

Recent development has used counterfactual result to decompose the target metric and reduce the variance \cite{deng2024metric}. Specifically counterfactual results are utilized to categorize the events into high and low signal-to-noise ratio portions and it enables the weighting between the two portions. Our work is built on top of the approach.

\section{Interleaving with Competitive Pair}
As outlined in Section ~\ref{sec:intro}, our existing experimentation process comprises two steps. Initially, experimenters utilize offline evaluation with historical data during the early stages of iteration. This approach is fast and cost-effective, requiring only 1-2 hours for Airbnb cases, though it lacks accuracy. Subsequently, the promising candidates identified in the first step proceed to A/B testing, which is constrained by limited bandwidth and requires weeks to complete. This bottleneck could be alleviated by introducing a middle stage that that is much faster than A/B and more accurate than offline evaluation. Interleaving, known for its high sensitivity \cite{chapelle2012large}, emerges as a viable option for the stage. We tackle the online evaluation challenges 1) low frequency transaction and 2) conversion as target metric by efficient team draft design and innovative credit attribution.

\subsection{Methodology}
In Section ~\ref{sec:bg_iterleaving}, we examined three variations of interleaving: BI, TD, and PI. The complexity of these methods, particularly in terms of their merging algorithms and credit attribution processes, can be ordered from most to least complex as PI > BI > TD. Given that our system is intended to operate in a production environment and cater to all search ranking experimentations, we chose TD for its efficiency and extensibility. 

\subsubsection{Competitive Pair and Credit Attribution}
In the previous work \cite{radlinski2008does}, the team drafting process involves teams taking turns to select the next item not yet included in the merged list $I$. A coin flip at each turn determines which team picks first. We design team drafting with a notion of competitive pair and the coin flip only happens once at the beginning procedure. In every turn, we draft the next available item from each team. If different, they form a competitive pair, and are added $I$ with the order determined by the coin flip. The team assignment is logged accordingly. If the items are identical, the item is added to $I$ without being assigned to any team. This procedure is detailed in Algorithm ~\ref{alg:tmc}. The design is highly efficient for serving and preference calculation. To illustrate, let's walk through an example with two ranked lists $C = \{a, b, c, d, e \}$ and $T = \{b, c, a, f, g \}$, assuming $isCfirst=true$. The first few steps of team drafting process would unfold as follows:
\begin{itemize}
\item Draft $a$ from C and $b$ from T. As the two items are different, they form a competitive pair and are placed in $I$, as  $[a^C, b^T]$(line 7 - 9). The super script indicates the team assignment.
\item The next available items are $c$ from $C$ and $c$ from $T$. Since they are the same, $c$ is added to $I$ without a team assignment.  (line 13 - 14)
\item This process continues until the end condition is met.
\end{itemize}
Following this procedure, the output would be $I = \{ a^C, b^T, c, d^C,  f^T \}$. Our design guarantees that each team has an equal opportunity to be selected first, ensuring that rankers C and T have identical chances of displaying their listings in any given position. The approach effectively eliminates position bias in our measurements. To maintain a consistent user experience, we construct list $I$ with a length equal to the minimum of $l_c$ and $l_t$, the lengths of the lists generated by rankers C and T, respectively. In our production environment, $l_c$ and $l_t$ are equal except in rare cases, as it is from the same user search request.
\setlength{\textfloatsep}{0pt}

\begin{algorithm}[h!]
\DontPrintSemicolon
\KwIn{ranked list $C$ and $T$, coin filp result $isCfirst$}
\KwOut{Interleaved list $I$}
$l_c, l_t \gets len($C$), len($T$)$\;
$l_I \gets min(l_c, l_t)$\;
$k_c, k_t \gets 0, 0$\;
$I\gets \emptyset$ \;
\While{$k_c != -1 \& k_t != -1 \& k_c < l_c \& k_t < l_t \& len(I) < l_I$} {
	\If {$A[k_c] != B[k_t]$} {
		\If {$isCfirst$} {
			$I \gets I \cup C[k_c]^C$\;
			 $I \gets I \cup T[k_t]^T$\;
		}
		\Else{
			$I \gets I \cup T[k_t]^T$\;
			$I \gets I \cup C[k_c]^C$\;
		}
	}
	\Else {
		$I = I \cup C[k_c]$\;
	}
	$k_c \gets nextAvailable(C, k_c, I)$\;
	$k_t \gets nextAvailable(T, k_t, I)$ \;
}
\Return {$I$}
\caption{Competitive Pair Team Drafting}\label{alg:tmc}
\end{algorithm}

\subsubsection{Preference}
Team preference is determined by counting the victories of each competitive pair and then aggregating these results to the desired level of analysis. In our case, we consider the individual user as the unit of analysis (although search requests could also serve this purpose). For each user $i$, the outcome is identified based on which team—C or T—has more wins. We define $Y_i(w) = wins(w)$, where $w$ can either be team C or team T and the function $wins(.)$ simply counts the number of wins for the corresponding team. The preference measure, $\tau_i$, is then calculated as the difference in wins between team T and team C, expressed as $\tau_i = Y_i(T) - Y_i(C)$. This method provides a clear and quantifiable metric for team preference at the user level.
\begin{equation}
\tau_{pref} = \frac{1}{N}(\sum_i\mathbbold{1}_(\tau_i > 0) - \sum_i\mathbbold{1}_(\tau_i < 0))
\end{equation}

The p-value is computed by proportional test on $\sum_i\mathbbold{1}_(\tau_i > 0)$ and $\sum_i\mathbbold{1}_(\tau_i < 0)$, which are the number of subjects who prefers T and C respectively. 

Using competitive pairs allows for a direct comparison between two items from the two rankers, with each pair acting as the basis for attributing credit. This approach minimizes noise in determining preferences, especially when the event is sparse, such as conversion. It has demonstrated high sensitivity in our validation process.

\subsubsection{Unbiasness}
If users interact with $I$ by clicking randomly, there is no team preference based on the credit attribution scheme. To illustrate it, we use competitive pairs as the base unit to derive the user preference. A user who take actions randomly will click first and second item in the pair with probability $P(C=1|r=1)$ and $P(C=1|r=2)$, where $r$ is rank of the listing and $C$ = 1 if clicked, and 0 otherwise. The expectation of the listing from A (or B) get clicked is  $P(C=1|r=1)P(r=1) + P(C=1|r=2) P(r=2)$.  As each ranker has the equal chance of being ranked at the first according to team draft algorithm, so we have $P(r=1) = P(r=2) = 0.5$. By aggregating all pairs, we will get that ranker A and B will have an equal number of expected wins.

Careful readers may question the unbiasness when only the first item from the last competitive pair is returned, given our control over result length by $l_I$ in Algorithm ~\ref{alg:tmc}. Since each team has an equal chance of having this lone item, no bias is introduced, as confirmed by the data quality checks detailed in the following section and Section ~\ref{sec:interleaving_unbiasness_validation}.

\subsubsection{Data Quality Monitoring}
\label{sec:interleaving_data_quality}
In addition to assessing the business impact of the treatment ranker, our credit assignment algorithm serves an important role in data quality checks, which are essential for accurate experimentation. Specifically, the algorithm can be applied to evaluate the distribution of impressions and the frequency with which each team appears first in a competitive pair. We anticipate neutral outcomes since each team should receive an equal number of impressions, and there should be a 50\% chance for either team to appear first in the competitive pair. The methodology enables us to verify the unbiased nature of each experiment. Should these quality metrics fail to meet our expectations, the results of the experiment would be deemed invalid, indicating that the team drafting algorithm did not perform as expected. To our knowledge, this approach has not been previously proposed.

\subsection{Architecture Design}
\label{interleaving_delivery}
\begin{figure}[h]
  \includegraphics[width=\linewidth]{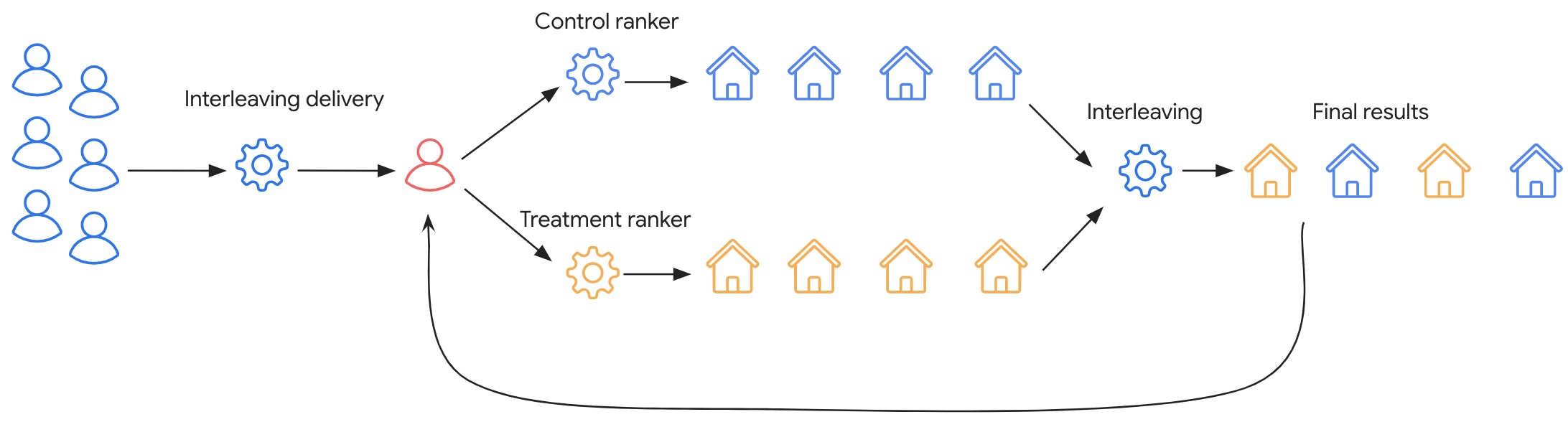}
  \caption{Interleaving delivery system. Twos layers of randomization are used. The first layer decides which user is subject to interleaving and the second assigns the user to the specific interleaving experiment.} \label{fig:delivery}
  \Description{Interleaving delivery system. Two layer of randomization is used. The first layer decides which user is subject to interleaving and the second assigns the user to the specific interleaving experiment.}
\end{figure}

In order to support the interleaving, we designed two-layer experiment delivery scheme. The first layer divides the traffic, which are users in our case, into regular A/B test and interleaving. The users assigned to A/B portion will be exposed to A/B tests as usual. For whose who are assigned to the interleaving portion, the second layer maps them to the corresponding interleaving experiment (Figure \ref{fig:delivery}). Within the interleaving framework, each experiment slot is referred to as a "lane," highlighting the fact that all necessary traffic for interleaving is contained within this slot, unlike in the A/B setup, which requires a control arm in addition.
When the search system receives a query from a user assigned to interleaving, a parallel call component initiates control and treatment search requests simultaneously. These requests proceed through the entire search stack, producing individual search responses. Subsequently, the team drafting algorithm is employed to merge these responses into a final result list, which is then presented to the user.

\subsection{Interleaving in Production}
The interleaving system has been integrated into Airbnb's search ranking experiment process and has been utilized to conduct over a hundred experiments. Due to its high sensitivity, interleaving is conducted with a small fraction of traffic and over a much shorter duration than A/B testing. The computational complexity, including latency, is well within the threshold required by the search system.

\section{Counterfactual Evaluation}
While interleaving is recognized for its high sensitivity, there are specific scenarios where its application is limited. First, when a ranker extensively uses set level optimization, such as improving diversity of the results, the interleaving would break the assumption. Second, when search list is also used for generating other results on the website, such as map view at Airbnb, there is risk to disrupt user experience. 
Lastly, it is not straightforward to implement semantically meaningful metric for continuous value based outcome such as revenue.  A promising approach to overcome these limitations while preserving the benefits of search-level pairwise comparison is to avoid result blending altogether. This can be achieved by creating counterfactual results within the A/B testing paradigm, thereby maintaining the same user experience as outside of evaluation while still allowing for effective comparison and evaluation.

\begin{figure}[h!]
  \includegraphics[width=0.4\textwidth]{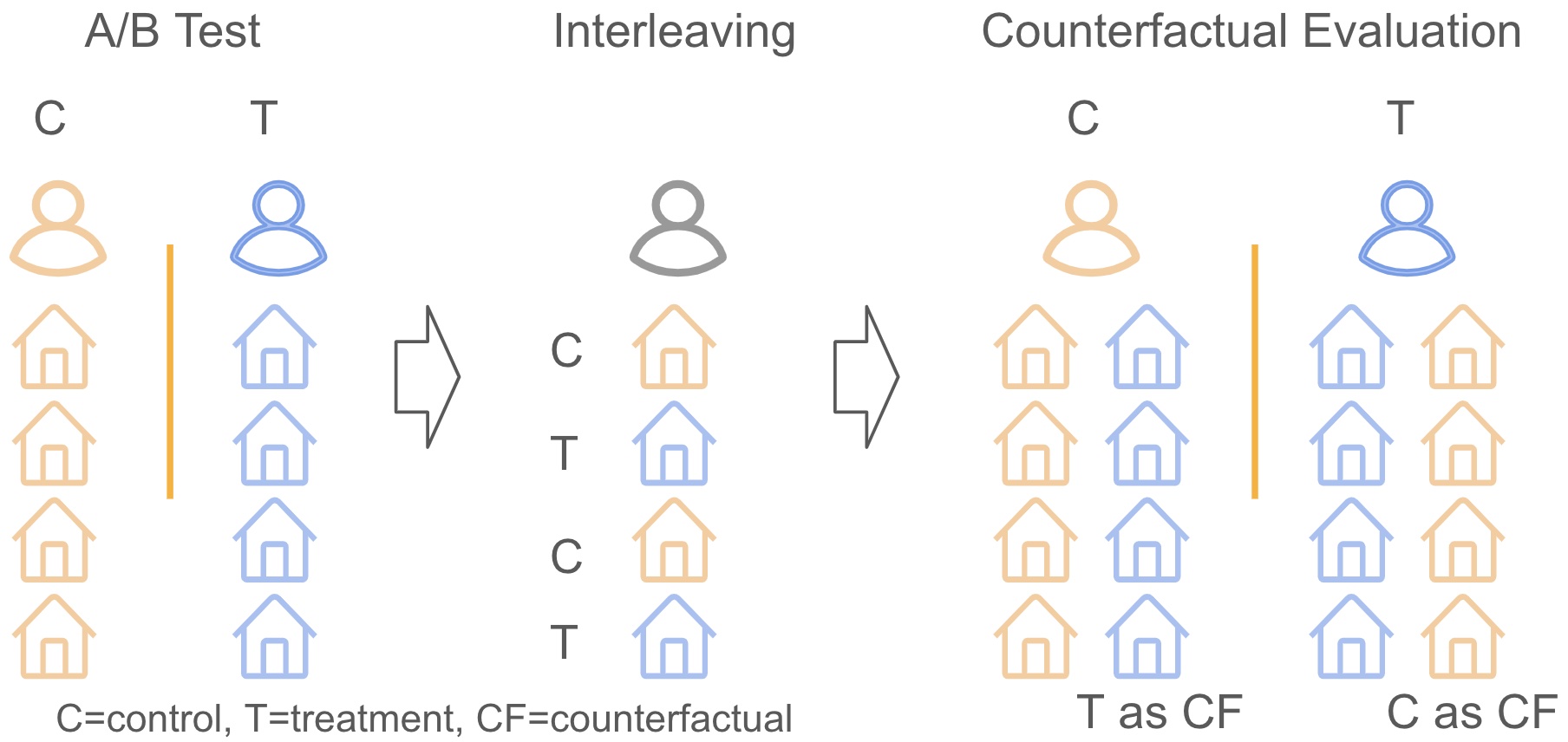}
  \caption{Counterfactual evaluation intuition. It can be understood as combining the characteristics of A/B tests and interleaving. It uses the results from both ranker for metrics computation, but doesn't need to blend them together.} \label{fig:cf_intuition}
  \Description{Counterfactual evaluation intuition. It can be understood as combining the characteristics of A/B tests and interleaving. It uses the results from both ranker for metrics computation, but doesn't need to blend them together}
\end{figure}

Next, let's delve deeper into the rationale behind the approach. As previously mentioned, in an A/B test, participants are divided into two groups. For those who are in control group, they will always see results from control ranker, and similarly treatment group user will see the treatment ranker results. Then we compare the conversion rate of treatment vs control group. Interleaving takes a nuanced approach to this comparison. For each search query, it generates results from both the control and treatment rankers, merges these results, and then displays the combined list to the user. Counterfactual evaluation serves as a hybrid of A/B testing and interleaving. It leverages the concept of generating paired results for each search, similar to interleaving, yet evaluates these results using metrics computed in a manner akin to A/B testing. The relationship between the three types of evaluations are illustrated in Figure \ref{fig:cf_intuition}. 

In counterfactual evaluation, there is notion of shown and counterfactual results. For each search, we generate results $L_c$ and $L_t$  based on both control and treatment ranker. If $L_c$ shown to the user, then $L_t$ is the counterfactual result, and the roles reverse if $L_t$ is shown instead. For the ease of description, we use $w$ denote the shown ranker and $1-w$ as counterfactual. 

For each user who are subject to the experiment, we flip a coin to decide which result to show. The randomization is seeded by user ID and experiment ID. Such design ensures the consistent user experience and minimize the carryover effect when running back to back experiments.

We present online counterfactual evaluation below, which is a direct extension of interleaving. It leverages parallel calls similar to interleaving to obtain the results from both control and treatment. Subsequently we utilize the counterfactual results to analyze the shown results, as opposite to interleaving which blends these two sets of results. This approach allows us to derive metrics that are more sensitive than those typically used in A/B testing. In  \cite{deng2024metric}, direct decomposition on target metrics based on counterfactual result is proposed. With it as foundation, we present a novel approach that's based on relative position and estimated outcome in the counterfactual, which is proved to be more sensitive.

\subsection{Direct Decomposition Based Estimators}
First we describe the Direct Decomposition approach \cite{deng2024metric}. Let $Y(w)$ denote the outcome of each group, where $w$ is the assignments, and its value 0 and 1 represent control and treatment respectively.  For any given item associated with an event (for example, a booking), its ranking position is denoted as $r_{i}(w)$ when subjected to the treatment $w$. We can categorize the outcome into two types,
\begin{equation}
\begin{aligned}
Y_{i}^{sim}(w) = \mathbbold{1}(r_i(w) <= k  \&  r_i(1-w) <= k \\ \&  |r_i(w) - r_i(1-w)| <= \alpha)
\end{aligned}
\end{equation}
\begin{equation}
Y_i^{diff}(w) = \mathbbold{1}(|r_i(w) - r_i(1-w)| >  \alpha )
\end{equation}
Where $k$ and $\alpha$ are hyperparameters encoding the mapping of ranking difference and true conversion impact. We simply use the values that's discussed in \cite{deng2024metric}, which are $k=4$ and $\alpha=2$. The direct decomposition based estimator is,
\begin{equation}
\tau_{decomp} = \tau_{diff} + \theta * \tau_{sim}
\end{equation}
$\tau_{diff}$ and $\tau_{sim}$ are the average treatment effect aggregated on user level  by following Equation ~\ref{eq:potential_outcome}.
As detailed in \cite{deng2024metric}, $\tau_{diff}$ has much smaller variance than $\tau_{sim}$, and significant variance reduction can be achieved from the re-weighting. We use $\theta=0.2$ in production.
\subsection{Estimated Reward Based Estimators}
\label{sec:reward_based}
The direct decomposition method categorizes target metrics based on the ranked positions generated by both the shown and counterfactual rankers. This approach utilizes estimators that are dependent on both the absolute and the relative positions of the item in question. Building on this, we introduce a novel type of estimators that focuses solely on the difference in positions.

Define $f: \mathbb{Z}^+ \to \mathbb{R}$ as a booking probability model. For a search $i$ that has booking event, let $Y_i(w) = 1$, then $Y_i(1-w) = f(r_i(1-w))/f(r_i(w))$, and the gain between shown and counterfactual is 
\begin{equation}
Y_i(w) - Y_i(1-w) =  1- f(r_i(1-w))/f(r_i(w))
\end{equation}
In our implementation, we chose to use exponential function for $f$, with $\gamma$ as decay factor, $f(r) = \gamma^{-r}$. It is based on the observation that the lower the listing is ranked, the less likely that the user is going to interact with it.

We also incorporate the notion of similar ranking proposed in direct decomposition, specifically when $|r_i(w) - r_i(1-w)| <= \alpha$ we consider the item is ranked at similar position between the shown and counterfactual. Formally, we compute the gain as
\begin{equation}
\label{eq: g}
g_i = 1 - \gamma^{max(|(r_i(w) - r_i(1-w)| - \alpha, 0))}
\end{equation}
Based on win/lose status, we have a pair of estimators 
\begin{itemize}
\item $\tau_i^{win}(w), \tau_i^{loss}(w) = g_i, 0$, if $r_i(1-w)  - r_i(w)  - \alpha > 0$  (If shown ranker $w$ ranks better, it gets $g_i$ as the gain at winning position)
\item $\tau_i^{win}(w), \tau_i^{loss}(w) = 0, g_i$, if $r_i(w) - r_i(1-w) - \alpha > 0$  (If counterfactual ranker $1-w$ ranks better, shown ranker gets $g_i$ as the gain at losing position)
\end{itemize}
When $|r_i(w) - r_i(1-w)|$ is within $\alpha$, item rankings are deemed similar, so $g=0$ accordingly. Regarding $\gamma$, a smaller value corresponds to a faster decay of attention curve and results in larger incremental gain. The estimator doesn't use the absolute position, but simply the difference. We count  difference of overall wins as gain estimator,
\begin{equation}
\begin{aligned}
\tau_g = \frac{1}{N}(\sum \mathbbold{1}(\tau_i^{win}(w=1) > 0) * \tau_i^{win}(w=1) \\- \sum \mathbbold{1}(\tau_i^{win}(w=0) > 0) * \tau_i^{win}(w=0))
\end{aligned}
\end{equation}

We also designed win-loss estimator,  which is to count for each treatment, the difference of the event that's in winning position and losing position.
\begin{equation}
\begin{aligned}
\tau_{win-loss} =\frac{1}{N} (\sum_i (\tau_i^{win}(w=1) - \tau_i^{loss}(w=1)) -\\  \sum_i (\tau_i^{win}(w=0) -  \tau_i^{loss}(w=0)))
\end{aligned}
\end{equation}

We aggregate the metrics across the users.
\subsection{OEC (Overall Evaluation Criteria) Metric}
We use a combination of direct decomposition and estimated reward based metrics to form the OEC metric (main metric) with the purpose of combining the potential benefit of both estimators. Specifically, we define
\begin{equation}
\tau_{oec} = \beta_1 * \tau_{decomp} + \beta_2 * \tau_g
\end{equation}
We currently simply assign equal weights, with $\beta_1= \beta_2=0.5$.

\subsection{Connection with Interleaving}
\label{sec:connection_with_interleaving}
In \cite{hofmann2013fidelity}, TD interleaving was pointed out to lack fidelity as well as sensitivity when there is a shift or insertion between two lists. For example, with $C = \{a, b, c, d \}$ and $T = \{b, c, d, a\}$, then $I=\{a^C, b^T, c, d\}$ (C first), or $I=\{b^T, a^C, c, d\}$ (T first). If the user booked listing $c$, we do not infer any preference as it is assigned to neither team. However, if we look at the original list, T is the better ranker because it ranks $c$ higher than C. As we've seen in counterfactual evaluation discussed earlier, shift-by-one is considered no gain when we set equal zone threshold $\alpha > 0$. In the validation Section ~\ref{sec:alpha_variation}, we show that $\alpha=2$ works better than even $\alpha=1$ (we did not collect data for $\alpha=0$), which supports the conclusion from interleaving that $C$ and $T$ is a tie if the booked item is $c$.

The findings of counterfactual evaluation deepened our understanding of the relationship between fidelity and sensitivity in interleaving. The fidelity violation actually doesn't impact TD's sensitivity.

\subsection{Event Attribution}
\label{sec:attribution}
Until now, our discussion of the event has been somewhat abstract. In this section, we aim to clarify how event is attributed. The click event serves as the most straightforward example, establishing a direct link between the search result and the user's action of clicking. However, our primary interest lies in conversion; therefore, booking is the key event we focus on.
The search journey of an Airbnb user often spans multiple sessions, meaning a listing that gets booked can appear in more than one searches. Insights from previous research on Airbnb's ranking system \cite{tan2023optimizing} suggest that a user's decision to book a listing is influenced by every instance the listing appears during their search journey, not merely the final one. Consequently, for both interleaving and counterfactual evaluation, we attribute the booking event to all occurrences where the listing was presented to the user during the experiment period, acknowledging the compound effect of repeated exposure on the booking decision.

\section{Validation and Analysis}
For both interleaving and counterfactual evaluations, we gather corresponding A/B test results, when available, to serve as ground truth for validation. Note that often times what eventually tested in A/B is different from interleaving/counterfactual evaluation, as the experimenter would do final adjustments based on the evaluation results. We only picked the cases with no or minor adjustment for validation.

Our aim is to achieve readings that are not only consistent with A/B test results but also demonstrate higher sensitivity. To assess the consistency between the proposed evaluation methods and A/B test outcomes, we focus on the correlation coefficient between point estimates from both evaluation approaches.
To this end, we compile a validation corpus comprising two lists of point estimates: one from our proposed evaluation method, denoted as $M^E$, and the other from A/B test results, denoted as $M^{A/B}$. For each ranker $i$, there exists a corresponding point estimate in both lists, obtained from the proposed evaluation method and the A/B test, respectively. The correlation coefficient between these two sets of point estimates is calculated using standard methods \cite{casella2024statistical}, providing a quantitative measure of the consistency between our proposed evaluation techniques and traditional A/B testing,
\begin{equation}
corr=\frac{Cov(M^{E}, M^{A/B})}{\sigma_{M^E} \sigma_{M^{A/B}}}
\end{equation}
Where $Cov(M^{E}, M^{A/B})$ is the covariance between $M^{E}$ and $ M^{A/B}$.  $\sigma_{M^E}$ and $\sigma_{M^{A/B}}$ are variance within each result list.

\subsection{Baselines}
In our study, we conducted a comparison of our methodologies against the two most relevant contemporary approaches in the field. Firstly, for the interleaving method, we benchmarked our speed improvements against the findings reported by \cite{bi2022debiased}. Like us, \cite{bi2022debiased} evaluated their interleaving approach in the context of its performance relative to traditional A/B testing. Secondly, for the counterfactual evaluation method, we compared our approach to \cite{deng2024metric},  because the work in is also aimed at search ranking evaluation, making it a directly comparable study to our own. Through these comparisons, we aim to highlight the advancements our approaches bring to the field.

\subsection{Interleaving} \label{sec:interleaving}
\subsubsection{Consistency with A/B} 
\begin{figure}[h!]
  \includegraphics[width=0.3\textwidth]{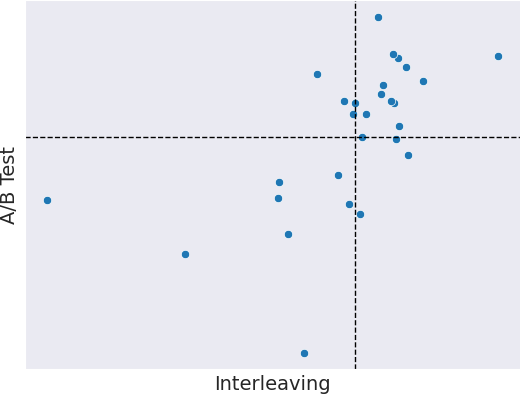} 
  \caption{Interleaving and A/B Test point estimates. Axis ticks are omitted.} \label{fig:interleaving_ab}
  \Description{Interleaving and A/B Test point estimates scattered plot. Axis ticks are omitted.}
\end{figure}

We collected 29 interleaving - A/B test pairs, and the point estimates of interleaving and A/B test on our target conversion metric is plotted in Figure \ref{fig:interleaving_ab}. Overall interleaving and A/B are directionally aligned 82\% of the time. The correlation coefficient is 0.6.

Through the usage, interleaving proved to be highly sensitive, as we observed about 50X speedup from A/B. We had a test ranker whose logic was to pick a random listing in the result list and put it to the top. The A/B test took weeks to conclude and the interleaving can detect the negative conversion impact using one day's data on a fraction of traffic. The BI based interleaving from \citep{bi2022debiased} reported 60X speedup based on a corpus size of 10. We consider the results are comparable and our approach is much more efficient computationally.
\subsubsection{Case Study}
We are particularly interested in the inconsistent pairs. Our cases suggested a limitation of interleaving when the set level optimization is involved. For example, there was a treatment ranker that optimizes another objective other than conversion. The aim was to remain neutral in terms of conversion while improving the secondary objective. Initially, listings were sorted according to their booking probability. The ranker then rearranged some of these listings to better align with the secondary objective. When users were directly presented with this re-ranked list, they were likely to book listings based on the estimated trade-off between the primary and secondary objectives.
However, the scenario changed when we interleaved these treatment results with control results. Users tended to select listings with a higher booking probability from the control group as they looked more attractive when placed side by side with treatment listings in competitive pair, leading to the treatment ranker's underperformance. This discrepancy was evident as we observed a significant negative impact in the interleaving results, whereas the conversion remained neutral in the A/B test. 

\subsubsection{Unbiasness Validation}
\label{sec:interleaving_unbiasness_validation}
As discussed in Section ~\ref{sec:interleaving_data_quality}, thanks to the extensibility of competitive pair based TD, we are able to compute data quality metrics in the same way as conversion. They provide the validation on unbiasness for each interleaving experiment and Table ~\ref{tab:interleaving_unbiasness} demonstrates the results from a past experiment. We expect no preference between two teams, which is confirmed by the metrics.

\begin{table}[!h]
\begin{center}
\caption{Unbiasness validation} \label{tab:interleaving_unbiasness}
\begin{tabular}{lcc}
\toprule
Metric & $\Delta$ & pval \\
\midrule
listings shown & 0.00\% & 0.91 \\
shown first & -0.01\% & 0.85 \\
shown reciprocal rank & -0.02\% & 0.85 \\
listings found & 0.00\% &  0.95 \\
\bottomrule
\end{tabular}
\end{center}
\end{table}

\subsection{Counterfactual Evaluation}
Similar to Interleaving, we collected 30 online counterfactual experiments whose treatments were later tested in A/B to evaluate the consistency and study the effect of hyperparameters. 

\subsubsection{Consistency with A/B}
\begin{figure}[h!]
\includegraphics[width=0.3\textwidth]{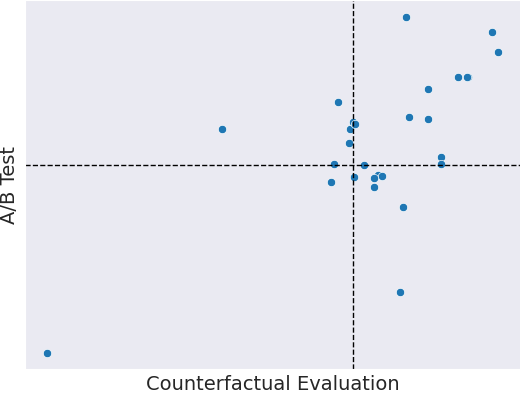} 
\caption{Online counterfactual evaluation and A/B point estimates. Axis ticks are omitted.} \label{fig:ab_vs_online_cf}
\Description{Online counterfactual evaluation and A/B point estimates scattered plot.}
\end{figure}

The main metric $\tau_{oec}$ point estimate, which is plotted in Figure \ref{fig:ab_vs_online_cf}, has correlation coefficient 0.65 with A/B. Therefore overall the consistency matches the interleaving.

We show individual metrics point estimate correlation with A/B tests in Table \ref{tab:metrics_vs_ab}. $\tau_g$ performs the best. $\tau_{diff}$ is much higher correlation than $\tau_{sim}$, which is consistent with findings in \cite{deng2024metric}. $\tau_{win-loss}$ is between the  $\tau_g$ and $\tau_{diff}$.

\begin{table}[!h]
\begin{center}
\caption{Metrics point estimate correlations with A/B test} \label{tab:metrics_vs_ab}
\begin{tabular}{lc}
\toprule
Metric & Corr with A/B \\
\midrule
$\tau_g$ & 0.66 \\
$\tau_{win-loss}$ & 0.58 \\
$\tau_{diff}$ & 0.55 \\
$\tau_{sim}$ & -0.29 \\
\bottomrule 
\end{tabular}
\end{center}
\end{table}

We studied the experiments in which the counterfactual evaluation didn't work well and they are mainly due to shared real time user signals. For each search, as shown and counterfactual results are derived from the same user, the user related features are identical for ranker $C$ and $T$. When two rankers have drastically different capability of utilizing such features, the stronger one would utilize the engagement earned by the weaker ranker and gain an unfair advantage. Therefore we may want to discount the trustworthiness when the control and treatment ranker pair falls into such case.
\subsubsection{Effect of Hyperparameters}
\label{sec:alpha_variation}
In reward based approach discussed in Section ~\ref{sec:reward_based}, there are two hyperparmeters. The first one is the position decay factor $\gamma$, quantifying how fast user's attention decrease along with the ranked position. We compared the $\tau_g$ and A/B correlation in respect to $\gamma$, and the results are similar. For $\gamma=0.9$ and $\gamma=0.95$, the correlation coefficients are 0.644 and 0.648 respectively.

The second hyperparameter we examine is ranked position similarity threshold $\alpha$, and our observations, as detailed in Table ~\ref{tab:alpha_variation}, indicate that setting $\alpha=2$ yields better correlations. This finding suggests that minor differences in relative position do not effectively differentiate ranking quality, regardless of the listing's overall rank.

\begin{table}[!h]
\begin{center}
\caption{Metric variations correlation with A/B ($\gamma=0.9$)}\label{tab:alpha_variation}
\begin{tabular}{lcc}
\toprule
Metric & $\alpha=1$ &  $\alpha=2$  \\
\midrule
$\tau_g$ & 0.64 & 0.66 \\
$\tau_{win-loss}$ & 0.58 & 0.60 \\
\bottomrule
\end{tabular}
\end{center}
\end{table}

\subsubsection{Sensitivity}
We observed significant speed up compared to A/B. To achieve our target minimal detectable effect, the counterfactual evaluation requires much less traffic. We observed around 15X speed up for $\tau_{oec}$, 23X for $\tau_{win-loss}$. The most significant speed up is from $\tau_{g}$, whose speedup is around 100X. The findings support our hypothesis that by adding additional knowledge from the counterfactual result of the same search, we can measure the impact with much less traffic.

\subsection{Interactions and Carryover Effects}
The risk of interaction between different evaluations is minimized through our experiment delivery strategy, which is outlined in Section ~\ref{interleaving_delivery}. The majority of search traffic is allocated to A/B testing, while the remainder is designated for interleaving or counterfactual evaluation methods. Within this smaller segment, we further divide the traffic into distinct lanes, with each experiment being assigned its own lane. This structured approach ensures that each evaluation method operates within its own controlled environment, significantly reducing the potential for cross-experiment interference.

The carryover effect refers to the influence of a previous experiment on the current one when they run back-to-back. Our approach minimizes this effect through randomization design. Let us consider an assignment group $G_i$, where $i \in \{C, T\}$ represents the control or treatment groups from the previous experiment, and let $u$ denote a user. In the case of interleaving, the carryover effect would occur if $\forall u \in G_i$, the condition $isCfirst$ is consistently true or false. However, this is avoided because randomization occurs at the search level - each search flips a coin to decide which ranker goes first in team drafting. It was further validated through experiments that the carryover effect wasn't observed.

For counterfactual evaluation, we employ randomization based on user ID and experiment ID. The latter ensures that the assignment for the current experiment is independent of the assignment for the previous experiment. Additionally, we have analyzed back-to-back experiments and found no evidence of carryover effects.
\section{Discussions}
\subsection{Implementation Choice}
Interleaving and counterfactual evaluation presents a promising direction of evaluation in ranking. The central idea is to have the visibility of the ranked lists from both ranker $\pi_1$ and $\pi_0$. For interleaving, the two lists are combined and shown to the user, so the speedup comes from the comparison in each competitive pair. The counterfactual evaluation does not interfere with what is going to be shown to the user, and simply uses the counterfactual results to create more sensitive reward estimators.

Interleaving and counterfactual evaluation metrics exhibit similar prediction power to the A/B tests, so generally speaking they both are well-suited for pre-A/B test online evaluations. Interleaving, with straightforward credit computation, has shown higher sensitivity compared to a subset of counterfactual evaluation metrics, which is a clear advantage.

On the other hand, counterfactual evaluation demonstrates greater robustness for rankers with strong set level optimization. In use cases like we discussed in the Section ~\ref{sec:interleaving} about re-ranking for optimizing secondary objective in Airbnb search, counterfactual evaluation would not suffer from the bias according to later experiments of similar nature, as user will always see the full results from control or treatment.

Therefore the choice of the technology is depending on the use cases,  as well as the experiment bandwidth availability.

\subsection{Generalization}
Both interleaving and counterfactual evaluation presented in the paper can be fairly easily applied to other businesses. They can be well applied to the scenarios when traffic (users) and the user action event are abundant, such as engagement-targeted (e.g. click through rate) optimization on search and recommendation. They would, in particular, show strength when traffic and/or events are limited, for instance, e-commerce platforms where the conversion is the target metric. Traditional A/B testing in such an environment demands prolonged periods, ranging from weeks to months, to gather sufficient data for reliable statistical power. Conversely, the methods we present require significantly less traffic and a shorter duration to yield meaningful results. We provide further guidelines on the implementation in the Appendix ~\ref{apdx:guideline}.

\section{Conclusion and Future Work}
The paper presented our innovation in speed up Airbnb search ranking experimentation. Our version of interleaving is efficient and highly sensitive, and we extended it to develop online counterfactual evaluation which addresses the limitations of interleaving and more generalizable. Both approaches are proved to be effective online evaluation technique for treatment candidate selection for A/B test based on the large scaled usage at Airbnb. The techniques can be easily adopted by other online platforms. 

Since implementation of these systems, we conducted hundreds of experiments for which we observed an increase in capacity to test new ideas and generally higher success rates in A/B testing. Furthermore, we leverage this framework for conducting model studies, including ablation tests and initial explorations for new projects.

For future directions, we aim to improve the accuracy of predicting outcomes from counterfactual results by incorporating data collected during the experiment itself. In our current approach in estimated reward based estimator, we rely on assumptions about outcomes associated with the counterfactual ranker. However, a more precise prediction model can be developed once data from the ongoing experiment, or on-policy data, is accessible. By analyzing user feedback from this data, we can refine our predictions, thereby enhancing the overall sensitivity of our evaluation method.

\bibliographystyle{ACM-Reference-Format}
\balance
\bibliography{citations}

\appendix
\section{Implementation Guidelines for Adoption}
\label{apdx:guideline}
We would like to provide suggestions on two areas that are key to adaption, which are event attribution and hyperparameter tuning.

\subsection{User Event attribution}
When applying our methods, practitioners need to carefully design the logic that attributes events, such as bookings, to the appearance of items, like listings shown in search or recommendation feeds. In cases where the booked listing appears in multiple search results, deciding on an attribution window becomes necessary. Options include attributing the booking to the appearance in the last search, to searches within the last two days, or to all searches within the experiment period. The choice could be based on the analysis on user decision making process.

\subsection{Counterfactual evaluation hyperparameters}
The selection of hyperparameters, specifically the attention decay factor $\gamma$ and the similarity threshold $\alpha$, is contingent upon the product's interface. For instance, a horizontal layout typical of recommendation systems may necessitate different parameter values compared to a vertical layout, which is common for search results. These parameters are crucial for accurately modeling user behavior and must be tailored to the specific characteristics of the product. Concretely, a possible procedure of tuning the parameters is as follows.

\begin{itemize}
\item Initial value $\gamma_0$ will be determined by curve fitting on the click or booking distribution across ranked positions, then we pick candidate values centered around $\gamma_0$. Subsequently we will mainly use meta-analysis to compare the metrics correlations that’s corresponding to each value with A/B tests.
\item $\alpha$ value will be determined by meta-analysis on values such as \{0, 1, 2, 3\}.
\end{itemize}

\end{document}